\documentclass{iaus}
\usepackage{graphicx}

\title[Composition of super-Earths]
{Composition of Transiting and Transiting-only super-Earths}
\author[Diana Valencia]
{Diana Valencia$^1*$}
\affiliation{$^1$ Universit\'e de Nice-Sophia Antipolis, CNRS UMR 6202, \\
              Observatoire de la C\^ote d'Azur, B.P. 4229, 06304 Nice Cedex 4, France \\ 
email: dianav@mit.edu  \\[\affilskip]
$*$ Now at 54-1710, Earth, Atmospheric and Planetary Sciences Department, \\
Massachusetts Institute of Technology, 77 Massachusetts Ave, Cambridge, MA, 02139 }

\pubyear{2011}
\volume{276}  
\pagerange{1--8}
\jname{The Astrophysics of Planetary Systems: Formation, Structure, and Dynamical Evolution}
\editors{A. Sozzetti, M. G. Lattanzi \& A. P. Boss, eds.}

\begin{document}

\maketitle

\begin{abstract}
The relatively recent detections of the first three transiting super-Earths mark the beginning of a
subfield within exoplanets that is both fruitful and challenging. The first step into characterizing these
objects is to infer their composition given the degenerate character of the problem. The calculations
show that Kepler-10b has a composition between an Earth-like and a Mercury-like (enriched in iron)
composition. In contrast, GJ 1214b is too large to be solid, and has to have a volatile envelope. Lastly,
while three of the four reported mass estimates of CoRoT-7b allow for a rocky composition, one forbids it
and can only be reconciled with significant amounts of water vapor. In addition to these three transiting
low-mass planets, there are now more than one thousand Kepler planets with only measured radius. Even
without a mass measurement (``transiting-only'') it is still possible to place constraints on the amount of
volatile content of the highly-irradiated planets, as their envelopes, if present, are flared. Using Kepler-9d
as an example, we estimate its water vapor, or hydrogen and helium content to be less than 50\% or 0.1\%
by mass respectively.

\keywords{super-Earth, CoRoT-7b, GJ~1214b, Kepler-10b, composition}
\end{abstract}
\firstsection

\section{Introduction}

In the quest for finding habitable worlds, the efforts to detect small
planets are starting to pay off with the discovery of the first three
transiting super-Earths: CoRoT-7b, GJ1214b and Kepler-10b. They exemplify
the fruitful results of three different missions that have the potential
to detect low-mass planets: the french-led CoRoT mission (\cite{COROT:2003}),
the MEarth ground mission that surveys the nearest M dwarfs stars
(\cite{MEarth:2009}) and the space mission Kepler, that can detect planets as small as Earth thanks to its unprecedented precision (\cite{Kepler-Mission}).
In the next few years, the count for super-Earths is expected to grow,
especially from objects being measured by Kepler. However, many of
them will not have measured masses as Kepler's targets are a challenge
for the radial velocity telescopes observing the same field of view.
We will have to wait for HARPS-NEF (\cite{Harps-nef}) to be built
before measuring their masses. 

Armed with masses and radii, and an appropriate internal structure
model we can infer the composition of super-Earths. Owing to the degenerate
character of the problem, there is no unique solution, and what we
can infer is bounds to the composition. However, the very short period
planets have an additional constraint, as their insolation values
are very high making their atmospheres susceptible to escape. This
is the case for the first three transiting planets, especially CoRoT-7b
and Kepler-10b that have orbital periods of less than one day! In addition,
there is one hot transiting-only planet, Kepler-9d, for which only
the radius is known. Despite the lack of information on its mass,
the planet is also so irradiated, that it is possible to place some
constraints on its composition. 

In these proceedings I will describe what we have learned about the
composition of each of the transiting low-mass planets, and use Kepler-9d
as an example to show what can be inferred for transiting-only hot
super-Earths.

\section{Model}

In contrast to the structure of the gaseous planets, which have very small cores compared
to their fluid H/He envelopes, super-Earths and to some extent mini-Neptunes
are mostly dominated by their rocky/icy cores, and their
atmospheres play only a small role in terms of their bulk composition.
This is why an internal structure model that takes into account the
complexity of rock and ice structure, as seen in the terrestrial planets
and icy satellites in our Solar System, is the appropriate one for
super-Earths. Despite having non-massive gaseous envelopes (up to several earth-masses), mini-Neptunes
call for a model that calculates correctly the temperature structure of the envelope, as it has a substantial effect on the density of water vapour and
H/He.

The results presented here were obtained by combining the internal
structure model by Valencia et al (\cite{Valencia_et_al:2006,Valencia_GJ876d})
for the rocky/icy interiors with the internal structure model CEPAM
(\cite{CEPAM:1995,Guillot:IntGiants}) for the gaseous envelope composed
of H$_{2}$O and/or H/He. While the former model has been applied
and tested for super-Earths and the solid planets in the solar system,
the latter has been used successfully to model the giant planets in
the solar system and in extra solar systems. The boundary condition
at the solid-gaseous interface satisfies continuity in pressure and
mass. \cite{Valencia:CoRoT7b} discuss details of the combined
model.

\section{Results}

\subsection{CoRoT-7b and Kepler-10b}

CoRoT-7b and Kepler-10b are planets that share similar features: their
radii, their period, the type of host star and perhaps their mass.
Their major differences are their age, and the uncertainty on
their masses. CoRoT-7 is a very active star, which makes the radial velocity data very noisy.
Several studies have suggested different values for the mass, with
a combined uncertainty of $1-10\, M_{E}$ (see Table 1). However,
three of the four studies suggest that the mass is compatible with
a rocky composition. 

Kepler-10b, on the other hand, has a well determined mass and an exquisitely well determined
radius $R=1.416_{-0.036}^{+0.033}R_{E}$ (\cite{Kepler-10b}),
thanks to the tight constraints on the star allowed by asteroseismology.
The object is compact enough that a rocky composition is inferred. 

The range of possible rocky compositions vary mostly due to the amount
of total iron the planet has, which is distributed mostly in the core
and some in the mantle, unless the planet is undifferentiated. To
span the range of rocky compositions we consider two unlikely extremes:
a pure iron planet, and a planet devoid of any iron (close to a Moon-like
composition); and two compositions present in the solar system: Earth-like
(33\% iron core, 67\% silicate mantle, with an iron to silicate ratio
of 2), and a composition enhanced in iron similar to that of Mercury
(63\% iron core, 37\% silicate mantle).

\begin{figure}
\begin{center}
\includegraphics[width=3.4in]{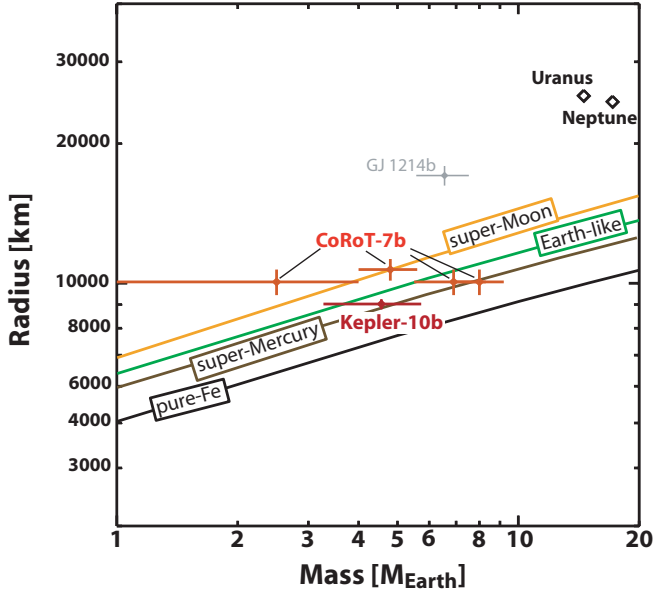}

\caption{CoRot-7b and Kepler-10b's composition as rocky planets. The relationships
for four rocky compositions: a super-moon (no iron), Earth-like (67\%
silicate mantle with 10\% iron by mol + 33\% iron core), super-Mercury
(in this study as 37\% silicate mantle with no iron by mol + 63\%
iron core), and pure iron are shown. Data for Kepler-10b, and CoRoT-7b 
(with its corresponding four mass estimates, and revised radius value -- see text for references) are shown. 
Uranus, Neptune and GJ~1214b are shown for reference}
\end{center}
\end{figure}

CoRoT-7b's first mass $M=4.8\pm0.8\, M_{E}$ (\cite{Queloz:CoRoT7b})
and radius $R=1.68\pm0.09\, R_{E}$ estimations (\cite{Leger:CoRoT-7b})
made it a planet lighter than Earth (see Figure 1). With improved
stellar parameters, the radius was revised to a smaller value of $R=1.58\pm0.10\, R_{E}$
(\cite{Bruntt:CoRoT7b}). This more compact scenario made the planet
more 'Earth-like'. Subsequently, the mass of the planet has been intensely
revised. While \cite{Hatzes:CoRoT7b} and \cite{Ferraz:CoRoT7b}
both obtain larger values of $M=6.9\pm1.4\, M_{E}$ and $M=8.0\pm1.2\, M_{E}$
respectively, making the planet denser and compatible with a composition
between Earth-like and Mercury-like, \cite{Pont:CoRoT7b} suggests
a mass of only $M=1-4\, M_{E}$ which can not be reconciled with a
rocky composition. This low value can only be satisfied with significant
amounts of volatiles.

On the other hand, the mass and radius of Kepler-10b is concordant
with a composition between an Earth-like and a Mercury-like planet.
Thus, if we disregard \cite{Pont:CoRoT7b} suggested mass value for
CoRoT-7b, both planets appear to be almost identical in composition.
This stresses the importance of establisthing the reliability of \cite{Pont:CoRoT7b}
treatment to infer the mass of CoRoT-7b. 

While we might be tempted at first to classify these planets as rocky
based on their bulk densities and proximity to their host stars, only
through a rigorous analysis on the likelihood of other compositions
can we be sure that they are indeed telluric planets. To this effect, we considered planets with different amounts of water vapor or a
hydrogen and helium mixture above an Earth-like nucleus, combined
with a simple analysis of atmospheric escape to determine if the timescales
of evaporation of these envelopes are consistent with the ages of
the planets (\cite{Valencia:CoRoT7b}). Figure 2 shows the results.
Based on a simple hydrodynamic espace treatment the timescale of evaporation
of water vapor is around a gigayear, while for H-He it is only a few
million years. Given the compact size of both planets, and the short
timescale of evaporation, we can infer that there is no H-He in both Kepler-10b
and CoRoT-7b, even with the low mass value suggested by \cite{Pont:CoRoT7b}.
It is also clear that Kepler-10b is too compact to allow for any significant
water vapor, plus the fact that it is a much older planet (at least
8 billion years old) suggests the evaporation would have taken place
for longer. Therefore, it is safe to assume it is a rocky planet. 

\begin{figure}
\begin{center}
\includegraphics[width=3.4in]{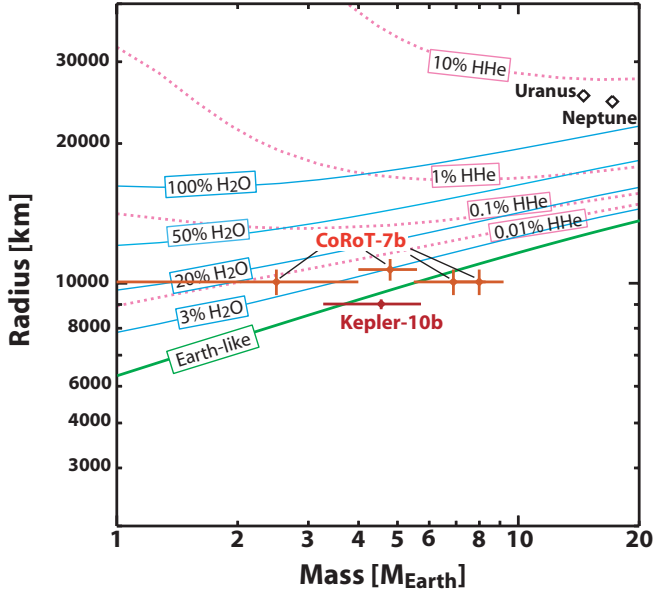}

\caption{Volatile content of CoRot-7b and Kepler-10b's. Data for CoRoT-7b and
Kepler-10b is shown. Two families of volatile content are shown: (in
pink) 0.01, 0.1, 1, and 10\% by mass of H-He and (in blue) 3, 20,
50, 100\% by mass of H$_{2}$O above an Earth-like nucleus. }
\end{center}
\end{figure}

For CoRoT-7b, the timescale of evaporation of water vapor is of the
same order of magnitude as the age of the system. Thus, it is possible
that water vapor has not evaporated completely from the planet. According
to \cite{Pont:CoRoT7b}'s estimate, CoRoT-7b has at most a few tens
of percents (less than 40\%) of water vapor (see Fig. 2). For the
larger mass values suggested by \cite{Queloz:CoRoT7b}, and \cite{Hatzes:CoRoT7b}
, there is less than a few percents of water vapor. And for the mass
values suggested by \cite{Ferraz:CoRoT7b} to have some water vapor,
the solid nucleus beneath has to be significantly enhanced in iron
with respect to an Earth-like composition. \\

\begin{table}
  \begin{center}
\caption{Data for transiting super-Earths and Kepler-9d}
 {\scriptsize
\begin{tabular}{|c|c|c|c|c|c|c|}\hline
\vspace{2mm}
 & {\bf Mass} & {\bf Radius} & {\bf Period} & {\bf Teff}  & {\bf Age} & Ref \\
 &  (M$_{E}$) &  (R$_{E}$) & (days) & (K)  & (Gy) &  \\ \hline
 
CoRoT-7b   &    & 1.68$\pm$0.09  & 0.854 & 1800-2550$^a$ & 1.2-2.3 & \cite{Leger:CoRoT-7b} \\
    &  4.8$\pm$0.8   &   &   &   &   & \cite{Queloz:CoRoT7b} \\
    &     & 1.58$\pm$0.10  &   &   &   & \cite{Bruntt:CoRoT7b} \\  
    &  6.9$\pm$1.4  &   &   &   &   & \cite{Hatzes:CoRoT7b} \\  
    &  8.0$\pm$1.2   &   &   &   &   & \cite{Ferraz:CoRoT7b} \\
    &  1-4   &   &   &   &   & \cite{Pont:CoRoT7b} \\ \hline 
Kepler-10b & $4.56_{-1.29}^{+1.17}$ & $1.416_{-0.036}^{+0.033}$ & 0.837 & 2150-3050$^a$ & 11.9$\pm$4.5 & \cite{Kepler-10b} \\ \hline
GJ 1214b   & 6.55$\pm$0.98 & 2.678$\pm$0.13 & 1.58 & 500-700 $^b$  & 3-10 & \cite{GJ1214b} \\ \hline 
Kepler-9d  & - &  & 1.59 & 1620-2300$^a$ & 2-4 & \cite{Holman:Kepler9}\\ 
           & - & $1.64_{-0.14}^{+0.19}$  &   &  or 1800-2500 $^b$ &   &  \cite{Torres:Kepler9d} \\ \hline
 \end{tabular}
 }
\end{center}

\vspace{1mm}
 \scriptsize{
 {\it Notes:}\\
  $^a$Effective temperatures are calculated with an albedo = 0 assuming a rocky composition, assuming no and full redistribution over the planetary surface. \\
  $^b$Effective temperature calculated with an albedo = 0.3 assuming a water composition, assuming no and full redistribution over the planetary surface.}
\end{table}

\subsection{GJ 1214b}

In contrast to CoRoT-7b and Kepler-10b, there is no ambiguity that
GJ 1214b has a volatile envelope. This planet has a radius that is
about 1000 km larger than if it was made of pure ice, the lightest
composition for a solid planet. The important question is the nature
of this envelope. As a starting point, we consider planets that have
different amounts of water vapor above an Earth-like nucleus, up to
a composition that is of pure H$_{2}$O (see Fig. 3). The boundary
condition we used is based on the calculations by \cite{Miller-Ricci:GJ1214b}
for the pressure-temperature of the atmosphere of this planet. For
different compositions, they obtain a value close to 1000 K at 10
bars. For this boundary condition, we find that GJ 1214b can be made
of 100\% water vapor. However, this composition is unlikely. Before
water can condense out of the solar nebula, refractory material has
to condense out. An upper limit to the ratio of water to refractory
material can be estimated from the composition of comets, with a dust
to gas ratio of 1-2. Therefore, given the fact that we expect some
rocky material to be present in the composition of this planet, we
also expect a component ligher than water. We propose this to be a
hydrogen and helium mixture (see \cite{Rogers_Seager:GJ1214b} for other
suggestions). We calculate the maximum amount of H-He to be 8\% by mass by adding varying
amounts of refractory material in the form of an Earth-like composition. An important note to make
is that this result hinges on the boundary condition for the atmosphere
that we chose. At this point, we are testing the sensitivity of this
result. 

\begin{figure}
\begin{center}
\includegraphics[width=3.4in]{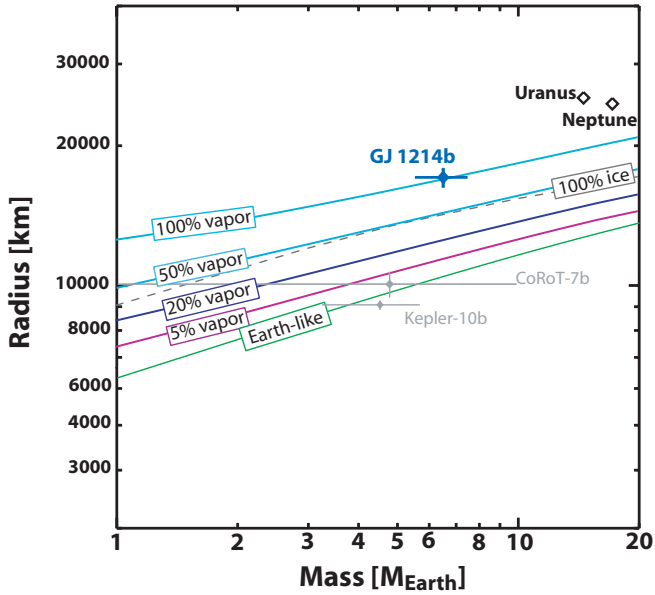}

\caption{GJ 1214b volatile composition. Data for CoRoT-7b and Kepler-10b is
shown. Mass-radius relationships for a composition of 5, 20, 50, 100\%
by mass of H$_{2}$O above an Earth-like nucleus. The pure-ice line
shows that GJ 1214b is a vapor planet. }
\end{center}
\end{figure}

One advantage of characterizing GJ1214b, is that the system is amenable
to further for observations. In fact, two different groups have obtained transmission
spectra to constrain the composition of the atmosphere at the millibar
lever. While \cite{Croll:GJ1214b} suggest that the composition is
H-He, \cite{Bean:GJ1214b} report a non-detection. To reconcile their
results, the latter authors argue that they might be seeing hazes. 

At first value, our result of an envelope with H-He making up to 8\%
by mass of the planet seems consistent with the results \cite{Croll:GJ1214b}.
If there is H-He at the 10 bar level and deeper, it is reasonable
to expect some of it also at the millibar level, more so than water. To properly
characterize GJ1214b two things have to happen: the observations have
to be reconciled in a coherent picture of what the composition is
at the millibar lever, and then this has to be reconciled with the
theory of the internal structure of the planet.

\subsection{Transit-only Planets: Kepler-9d}

Early in 2011, the Kepler team announced more than 1000 planet candidates, most of them having only measured radii (\cite{KeplerMission-2011}).
Even though there cannot be any absolute inferences on the composition
of a planet without a mass measurement, it is still possible to put
limits on the amount of volatiles of the short-period planets that
only have a radius measurement. An example is Kepler-9d with a measured
radius of $1.64{}_{-0.14}^{+0.19}\,\, R_{E}$ (\cite{Torres:Kepler9d}).
This is the smallest object orbiting Kepler-9 at 1.59 days, in a system
with two saturn-like planets in close resonance at 19.2 and 38.9 day
periods (\cite{Holman:Kepler9}). Even though there is no confirmation
from radial velocity observations, (\cite{Torres:Kepler9d}) have done
a careful analysis to assess the likelihood of false positives from
which they conclude that Kepler-9d is most likely a planet. A maximum
mass of 15-20 $M_{E}$ is estimated from a non-detection in the radial
velocity measurements (private communication with M. Holman).

Figure 4 shows the radius range for Kepler-9d and different mass-radius
relationships. One advantage of the short-period planets is that if
they have envelopes they are very hot and expanded, limiting the amount
of volatiles present in the composition for a given radius constraint. This is exemplified by the
flaring of the mass-radius curves for H-He compositions for planets
with masses below 5 $M_{E}$ and to some extent of the H$_{2}$O compositions
for planets with masses below 2 $M_{E}$. These planets do not have
enough gravity to keep tightly bounded their very hot and expansive
envelopes. This means that for Kepler-9d we can rule out compositions
with more than a few tens in 10000 parts of H-He by mass, and more
than 50\% water vapor content. 

On the other hand, for this planet to be rocky, its mass would have
to be more than 3 $M_{E}$. Larger masses would call for more
iron content, with an Earth-like composition corresponding to a mass
range between 4-10 $M_{E}$, a Mercury-like composition to the range 7-15
$M_{E}$, and a pure iron composition to the range 13-30 $M_{E}$. The
latter would be an unlikely composition, perhaps possible only after complete
evaporation of a silicate mantle. Thus, based on physical grounds,
the absolute maximum mass for Kepler-9d is 30 $M_{E}$, although a
more realistic upper limit is $\sim$15 $ $$M_{E}$, when using
Mercury's composition as a proxy. 

\begin{figure}
\begin{center}
\includegraphics[width=3.4in]{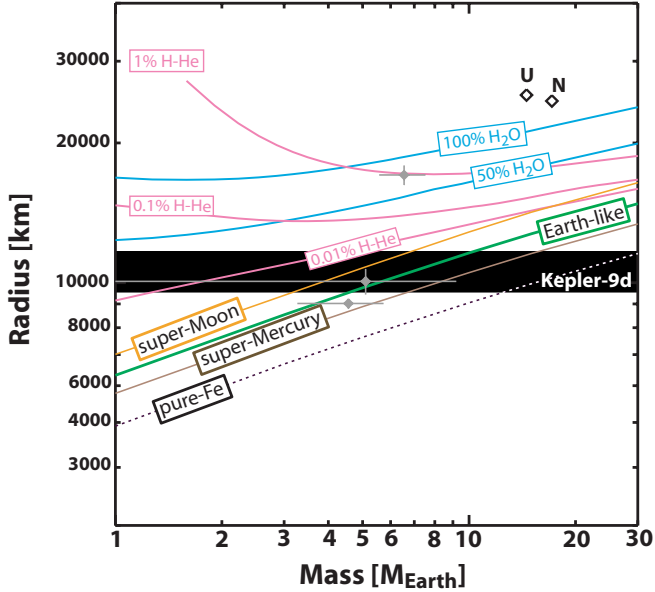}

\caption{Kepler-9d's composition. Data for CoRoT-7b and Kepler-10b is shown.
Two families of volatile content are shown: (in pink) 0.01, 0.1, and
1\% by mass of H-He and (in blue) 50, 100\% by mass of H$_{2}$O above
an Earth-like nucleus. Four rocky compositions are shown varying in
the amount of iron content: super-moon (no iron), Earth-like, Mercury-like,
to pure-iron planet (see figure 1 for detailed description of compositions).
The radius data for Kepler-9d is shown in black. In grey, the data
for Kepler-10b, CoRoT-7b and GJ1214b are reproduced for reference. See also related work in \cite{Havel:Kepler9}. }

\end{center}

\end{figure}

\section{Conclusions}

Now that we have formally entered the era of super-Earths with the
first transiting such planets having measured masses and radii, we can
start placing constraints on their composition. Given the degenerate
character of the problem, the first challenge is to differentiate
the planets that are mainly solid (rocky or icy) from the ones that have
non-negligible volatile envelopes (more than a few percent by mass).
This task plus the bias towards detecting short-period planets brings
to light the importance of better understanding and modelling atmospheric
escape, as these planets are highly irradiated by their host stars.
Any composition that allows for a volatile envelope has to be reconciled
with the planet's history of mass loss. 

From this initial small sample of low-mass planets, there are two
robust although simple conclusions: Kepler-10b is a rocky planet,
while GJ 1214b has a significant amount of volatiles (better termed
a mini-Neptune). Kepler-10b's composition ranges in iron content from
an Earth-like (67\% silicate mantle + 33\% iron core) to a Mercury-like
(63\% silicate mantle + 37\% iron core) composition. 

Although it is clear that GJ 1214b has a volatile envelope, it is
unclear what the nature of of this envelope is, and especially if it is water vapor
dominated or H-He dominated. From internal structure models, the maximum
amount of H-He possible is 8\% by mass. In addition, the two reported
observations on the scale height, and therefore the composition of
the atmosphere (at the millibar level) seem contradicting. We stand
to learn a lot about the mini-Neptunes through a careful characterization
of GJ 1214b. 

On the other hand, the nature of CoRoT-7b is controversial, given
the large uncertainty on its mass. Three of the four studies point
towards a plausible rocky composition, and one forbids it. Determining
if this planet has a mass less than 4 $M_{E}$ is key, to classify
it as a super-Earth or a mini-Neptune, with implications about its
origin and evolution. In any case, we can rule H-He in the envelope
given the relatively small size of the planet, and the short timescale of
evaporation of such an envelope compared to the age of the system. 

Lastly, even though it is impossible to characterize transit-only
planets, we can still estimate the maximum mass of these planets and put constraints
on their amount of volatile content (especially for the hot ones). One
example is Kepler-9d, for which we find that a reasonable maximum
mass based on physical grounds is 15 $M_{E}$ (consistent with a non-detection
given the precision of the radial velocity analysis), a maximum amount
of water vapor at the 50\% by mass level, and less than 0.1\% of H-He.

\end{document}